\documentclass[epj]{svjour}
\usepackage{graphics,amsmath,graphicx,float,color,ulem}
\newcommand{\kf}{\kappa_f}
\newcommand{\kx}{k_\times}
\newcommand{\ath}{\langle \theta\rangle}

\begin{document}

\title{Elasto-plastic response of reversibly crosslinked biopolymer bundles}
\titlerunning{Elasto-plastic biopolymer bundles}

\author{Poulomi Sadhukhan, Ole Schuman \and Claus
Heussinger
}                     
\mail{sadhukhan@theorie.physik.uni-goettingen.de,
heussinger@theorie.physik.uni-goettingen.de}

%
\institute{Institute for Theoretical Physics, Georg-August
University of G\"ottingen, Friedrich-Hund Platz 1, 37077 G\"ottingen,
Germany }
%
\abstract{We study the response of F-actin bundles to driving forces
  through a simple analytical model. We consider two filaments
  connected by reversibly bound crosslinks and driven by an external
  force. Two failure modes under load can be defined. \textit{Brittle
    failure} is observed when crosslinks suddenly and collectively
  unbind, leading to catastrophic loss of bundle integrity. During
  \textit{ductile failure}, on the other hand, bundle integrity is
  maintained, however at the cost of crosslink reorganization and
  defect formation. We present phase diagrams for the onset of
  failure, highlighting the importance of the crosslink stiffness for
  these processes. Crossing the phase boundaries, force-deflection
  curves display (frequency-dependent) hysteresis loops, reflecting
  the first-order character of the failure processes. We evidence how
  the introduction of defects can lead to complex elasto-plastic
  relaxation processes, once the force is switched off. Depending on,
  both, the time-scale for defect motion as well as the crosslink
  stiffness, bundles can remain in a quasi-permanent plastically
  deformed state for a very long time.
\PACS{
      {87.16.Ka}{Filaments, microtubules, their networks, and supramolecular assemblies}   \and
      {62.20.F-}{Deformation and plasticity}
      {87.15.La}{Mechanical properties}
{87.19.rd }{Elastic properties}
     } 
} 
\maketitle
\section{Introduction}
\label{intro}
Actin is one of the the most prominent examples of a protein that can
polymerize into long filamentous polymers (f-actin). In combination
with some of the many different actin-binding proteins
(``crosslinks'') these filaments can then assemble into a wealth of
higher-order cytoskeletal structures, with a multitude of different
biological functions. On a fundamental level it is an interplay
between the energy scales of crosslink and filament deformations that
determines the mechanical properties of these structures. The
dynamical properties are governed, among other factors, by the
reversibility of the filament-crosslink bond, which has an intrinsic
and finite lifetime. In living cells, dynamic crosslinking is needed
to facilitate the rearrangement of the cytoskeleton under external
mechanical and chemical forces. This results into dissipation of
memory of initial states and stabilization in new configurations under
changing environment. Under load, reversible crosslinking can be one
pathway for stress release. Consequences may be internal
rearrangements, large-scale structure formation, creep or even
catastrophic failure.

Apart from complex cytoskeletal networks, actin filaments are also
found to assemble into simple structures with only few filaments. For
example one can observe kinked helices, rings, tennis-racket shapes
due to a competition between elastic and interfacial
effects~\cite{cohenpnas,ceberprl}. The coiled acrosome is composed of
straight sections of bundled actin filaments joined via
kinks~\cite{derojmb}. A racket shape is also observed in a tubulin rod
that had buckled inside of a vesicle~\cite{fygprl}.

Here, we consider the arguably simplest filament assembly: that of two
filaments crosslinked together in a parallel fashion to form a bundle.

F-actin bundles are plenty in cytoskeletal structures in
eukaryotes. They provide mechanical stability in filopodia,
microvilli, stereocilia stress fibers and the sperm acrosome, play
roles in various cellular functions like
locomotion~\cite{mogbioj,atibioj,vigjcb},
mechanotransduction~\cite{hudpnas} and fertilization. Some in-vitro
experimental studies reveal that the crosslinking proteins and their
interactions with the filaments have important effects on the
mechanical and the structural properties of the bundle
assemblies~\cite{clanatmat,shinjmb,shinpnas,claessenspnas,purdyprl,shinprl,havivebj}. Under
the assumption of permanently bound crosslinks, a theoretical
description has been developed (``wormlike bundle'' model) to
characterize bundle mechanics~\cite{batbioj,heussingerPRL2007WLB}.
Theoretical work has also focused on thermal
denaturation~\cite{benpre,kieprl1}, thermally assisted force-induced
desorption~\cite{kieprl2}, or the effects of filament helicity on
bundle structure and stability~\cite{grasonPRE2009,heussingerJCP2011}.

In recent experiments~\cite{strehle}, the time dependent effects of
large bundle deformation has been studied in-vitro. These experiments
highlight the important aspect of ``crosslink remodelling'', which has
not been accounted for in these previous studies: in response to
force, crosslinks will repeatedly un- and rebind at different binding
sites along the actin filament. In the experiments, a crosslinked
f-actin bundle is subjected to a stress and kept in a bent state for a
short period ($10$s) or a long period ($1000$s). After this waiting
time, the bundle was released and the relaxation was observed with
time. For short waiting time, the bundle relaxes back exponentially to
its initial straight ground state. It behaves elastically. For the
long waiting times, however, the bundle only partially relaxes back
and a substantial residual bending deformation remained in the bundle,
which is therefore plastically deformed. It seems that upon
deformation, new crosslink binding sites become available which allow
to reduce the strain on the crosslinks and thus are more
favourable. After release of the force, these new connections
stabilize the bent conformation and thus the bundle remains in a bent
form.

This transition from elastic to plastic response of the filament
bundle is the subject of this paper. We will show how already a simple
two-filament bundle can display complex mechanical properties. With
the help of case-studies we can define two failure modes of the bundle
under load. \textit{Brittle failure} is observed when crosslinks unbind
under load~\cite{chpre,vinkjcp}, leading to catastrophic loss of
bundle integrity. \textit{Ductile failure} maintains bundle integrity at
the cost of crosslink remodelling and defect formation.

The paper is organized as follows. In Sect.\ref{sec:model}, we
describe our model. Two different cases of crosslink binding are
considered. The results are discussed in Sect.\ref{sec:results}. We
show that bundle deformation leads to crosslink unbinding processes
that crucially depend on the stiffness of the crosslinking protein.
The Sect. \ref{sec:unbind} and Sect.\ref{sec:rebind} discuss the two
above mentioned scenarios. A time-dependent force is introduced in
Sect.\ref{sec:dyn}. There we show how the response changes with the
frequency of the driving force, and also consider the relaxation of
the bundle once the force is switched off. Finally, in
Sect.\ref{sec:summary}, we summarize our results.

\section{Model}
\label{sec:model}

\begin{figure}[h]
  \centering
   \includegraphics[width=5.5cm,clip]{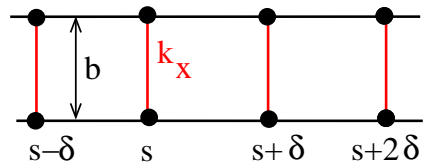}   
   \caption{Schematic diagram of an F-actin bundle with two
     filaments. Two black lines represent the filaments.
     The black circles show the crosslink binding sites on the
     filaments. The red vertical lines represent the crosslinks
     connecting two filaments. The distance between two crosslink
     binding sites is $\delta$ and the lateral distance between the
     filaments is $b$. The crosslink stiffness is denoted by $\kx$. }
  \label{fig:schem}
\end{figure}

We consider a biopolymer bundle of length $L$, lying in a
two-dimensional plane and consisting of two parallel inextensible
actin filaments. A schematic diagram of such a bundle is shown in
Fig.\ref{fig:schem}. In a variety of systems such as organic and
inorganic nanotubes as well as stiff biopolymers, the stretching
deformation mode is energetically expensive relative to the bending
mode, so that the filaments may be approximated as inextensible. We
comment on the effect of filament stretching at the end of
Sect. \ref{sec:results}. The filaments are laterally interconnected by
reversible crosslinks (red lines in Fig.\ref{fig:schem}) they can
dynamically bind and unbind the filament pair.

\begin{figure}[h]
  \centering
   \includegraphics[width=5.5cm,clip]{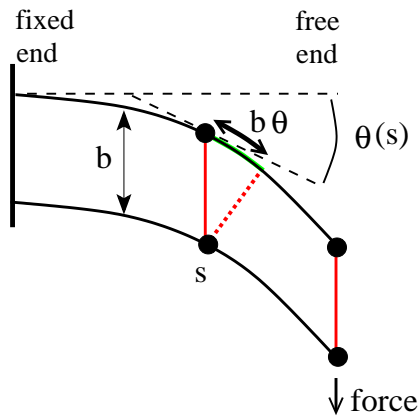}   
   \caption{A bent configuration of a crosslinked filament
     bundle. Bending by an angle $\theta$ leads to crosslink sliding
     and a shear deformation of amplitude $b\theta$.}
\label{fig:bentfig}
\end{figure}

On each filament there are $N_\times$ crosslink binding sites, spaced
at regular intervals a distance $\delta$ apart. One end of the bundle
is grafted at a wall and the other end is free to move. The force is
applied at the free end which produces a bending deformation in the
bundle creating stress on the crosslinks. The boundary conditions
resemble the in vitro experiments done by D. Strehle et. al
\cite{strehle}, where the free end is subjected to a force by pulling
it while the other end is immobilized by sticking to a heavy bead.
Because of the driving force, the bundle will be deformed by bending
the filaments and shearing the crosslinks (see
Fig.\ref{fig:bentfig}). The shearing
energy~\cite{everaers95,camalet1999PRL,hilfinger2008PhysBiol,heussingerPRE2010}
of the crosslinks amounts to
\begin{equation}
  \label{eq:Hsh}
  H_{sh}=\frac{\kx}{2\delta}\int_0^L \left(b\theta(s)\right)^2\, ds,
\end{equation}
where $b$ is the separation between the two filaments and $\kx$ is the
crosslink stiffness, or the mechanical stiffness of the crosslinking
agent. Here $\theta(s)$ is the angle of inclination, i.e. angle of the
local tangent the bent bundle makes at a point $s$ with respect to the
initial configuration, $s$ being the arc length along the bundle. The
quantity $b\theta(s)$, therefore, gives the amount of shear in the
crosslink at $s$.
The bending energy of the filaments can be obtained
from WLC Hamiltonian
as
\begin{equation}
  \label{eq:Hb}
  H_b=\kf\int_0^L \theta'^2(s)\, ds,
\end{equation}
where $\kf$ is the bending stiffness of the filaments. There is
another contribution coming in the Hamiltonian due the driving force
$f$ at the free end, which reads,
\begin{equation}
  \label{eq:Hf}
  H_f=-2f\int_0^L \theta(s)\, ds.
\end{equation}
In our calculations we use a linearized forcing term $\sim f\theta$,
which is only accurate for small deflections. For typical experiments
large deflections may be important, and the full expression
$f\sin(\theta)$ would have to be used. We have checked in some cases
that this would not change the qualitative behaviour of the bundle
response under stress. Moreover, considering the small $\theta$
approximation keeps the model solvable analytically.

The total Hamiltonian of the system is then the sum of the three
contributions, $H=H_{sh}+H_b+H_f$.

The most favourable configuration is chosen by the system by
minimizing energy. In doing so, some crosslinks may unbind from the
binding sites to cope with the force. This results in a change in the
crosslink density in the bundle depending on the magnitude of the
applied force, the stiffness of the crosslinks and bending stiffness
of the bundle. To proceed with the analytical calculations, we
introduce the crosslink density,
\begin{equation}
    \label{eq:definen}
    n={\rm number~ of~ bound~ crosslinks}/N_\times \nonumber
\end{equation}
in a mean field way which effectively normalizes the crosslink
stiffness $\kx\to n\kx$ in Eq.\eqref{eq:Hf}~\cite{vinkjcp,chpre}. The
crosslink density $n$ can vary from $0$ (all crosslinks unbound) to
$1$ (all crosslinks bound). There are two other contributions in the
energy, one is from the entropy of mixing, $E_{mix}=k_BT[n\ln
  n+(1-n)\ln (1-n)]$ and the other is $\mu n$, where $\mu$ is the
chemical potential for crosslink binding.

\begin{figure}[h]
  \centering
   \includegraphics[width=5.5cm,clip]{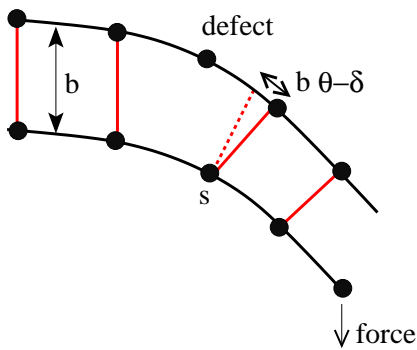}   
   \caption{A bent configuration of a forced crosslinked filament
     bundle. Rebinding of crosslinks has taken place to new binding
     sites after holding the bundle in bent configuration. The red
     dotted line shows the crosslink in undeformed state. The
     rebinding creates a defect in the bundle.}
\label{fig:bent}
\end{figure}

We modify the problem of unbinding of crosslinks in a way to allow the
crosslinks to rebind to new crosslink binding sites. To simplify
things, let us allow the crosslinks to rebind only to the right
neighbouring site, i.e. allow connections between site $\alpha$ of 1st
filament and sites $\alpha$ as well as $\alpha+1$ on the 2nd filament
(see Fig.\ref{fig:bent}).
The rebinding will result in a mismatch in the crosslink binding
registry between the two filaments. The starting point of such
mismatch will look like a ``defect'' with one vacant binding site in
one filament (Fig.\ref{fig:bent}b). We will only allow for one defect
in the following. If $s_D$ is the defect site, all the crosslinks
situated at $s>s_D$ will be rebound to the shifted registry. This can
be incorporated into the Hamiltonian by modifying the shearing energy
as,
\begin{equation}
  \label{eq:Hshd}
  H_{sh}=\frac{\kx}{2\delta}\int_0^L \left(b\theta(s)-d(s)\right)^2\, ds,
\end{equation}
where $d(s)$ is the relative shift in the crosslinking sites in the
two filaments. It goes from zero to $\delta$ when going through the
defect. For convenience we choose the following continuously differentiable
form
\begin{equation}
  \label{eq:ds}
  d(s)=\begin{cases}
0, &  \ s\leq s_D-\delta,\\
\frac{\delta}{2}\left(1+\sin\frac{\pi(s-s_D)}{2\delta}\right), & \ s_D-\delta<s<s_D+\delta,\\
\delta, &  \ s\geq s_D+\delta.
\end{cases}
\end{equation}
Here we assume that the defect has a core region of size $2\delta$,
from $s_D-\delta$ to $s_D+\delta$. Note that here we set $n=1$ as we
do not allow any open crosslink.

The total Hamiltonian is minimized with respect to $\theta$ to find
the corresponding differential equation. The solution for $\theta(s)$
is then plugged back into the Hamiltonian to find the energy. The
numerical calculations have been done with the following
parameters. The intercrosslink spacing $\delta$ and the separation
between the filaments $b$ are taken as $b=\delta=1$. In actin
filaments the crosslink sites can be taken to be roughly $40nm$ apart,
which corrresponds to the helical repeat of the
filament~\cite{clanatmat}. The bending stiffness $\kf=1$, sets the
energy scale in all the results. The length of the bundle and the
persistence length are taken as $L=l_p=500$, which correspond to $\sim
20\mu m$~\cite{kasn}. Temperature is fixed via
$(k_BT)^{-1}=l_p/\kf=500$. The most important parameter in this work
is the ratio $\kx/\kf$. We refer to this ratio whenever we say $\kx$
in this paper. Stiff crosslinks corresponds to large $\kx$ and soft
crosslinks corresponds to small $\kx$.  The force is measured with
respect to the Euler buckling force $f_b=\frac{\pi^2\kf}{2L^2}$.

\section{Results}\label{sec:results}
\subsection{Unbinding of crosslinks under force}\label{sec:unbind}

In this sub-section we consider the effects of unbinding under
increasing levels of force. To this end, we set $\kx\to n\kx$ in
Eq.~(\ref{eq:Hsh}) and determine the equilibrium state by minimization
with respect to $\theta(s)$ and $n$. Defect formation will not be
allowed in this section, thus we take $d(s)\equiv 0$.

The minimization with respect to the variable $\theta(s)$ gives a
differential equation, namely,
\begin{equation}
\theta''(s)-nK^2\theta(s)=-\frac{f}{\kf}\qquad K^2=\frac{\kx b^2}{2\kf\delta} \label{eq:diffeqt} ,
\end{equation}
solving which we get the behaviour of $\theta(s)$. The bending angle
satisfies the proper boundary conditions that $\theta(0)=0$,
$\theta'(L)=0$ and that $\theta(s)$, $\theta'(s)$ are continuous
everywhere. The bending angle is $\theta(s)\propto f$, more
explicitly,
\begin{equation}
  \label{eq:tksinf}
 \theta(s)=\frac{f}{K^2\kf}\left[1-\frac{\cosh K(L-s)}{\cosh KL}\right].
\end{equation}
Therefore, the effective free energy per
crosslink site including entropy of mixing is,
\begin{eqnarray}
  \label{eq:Etotksinf}
  F(n)&=&\frac{f^2\,\text{sech}^2 (\sqrt{n}KL)}{8n^{3/2}K^3\kf}\left[(10-3n)\sinh (2\sqrt{n}KL)\right.\nonumber\\
&&\left.-2\sqrt{n}KL(6-2n+(4-n)\cosh (2\sqrt{n}KL))\right]\nonumber\\
&&+ \mu\,n + k_BT[n\log n+(1-n)\log(1-n)].
\end{eqnarray}

\begin{figure}[htbp]
  \centering
  \includegraphics[width=7cm,clip]{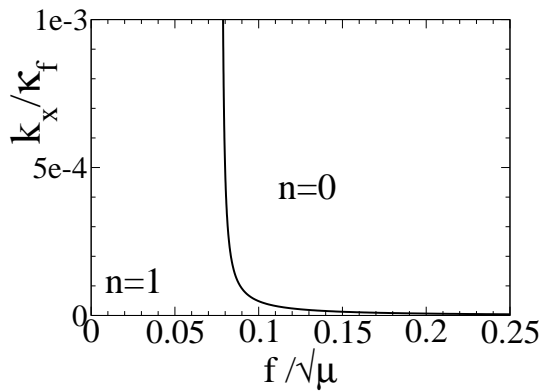}  
  \caption{Phase diagram for inextensible filaments at temperature
    $T=0$. The transition across the line separating the phases with
    average crosslink density $n=0$ (fully decoupled) and $n=1$ (fully
    coupled) is discontinuous.}
  \label{fig:phasediaksinfm}
\end{figure}

At zero temperature, the free energy has minimum either at $n=0$ or
$n=1$. The crosslink density corresponding to the lowest energy jumps
from $1$ to $0$ as we reach a critical value of force $f_c$. This
means that at $f_c$, suddenly all crosslinks unbind and the bundle
breaks down into two independent filaments. The phase diagram can be
found by equating $F(0)=F(1)$ (see Fig.\ref{fig:phasediaksinfm}).

\begin{figure}[htbp]
  \centering
  \includegraphics[width=7cm,clip]{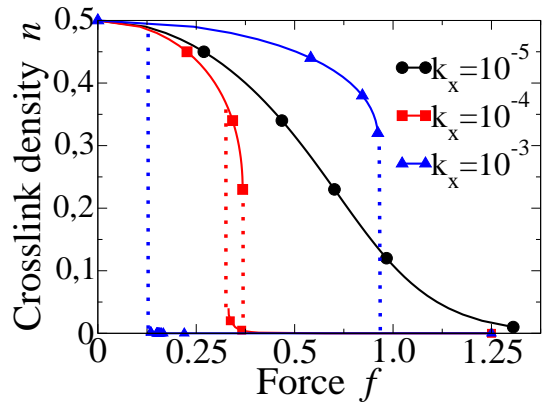}
  \caption{The variation of the crosslink density with driving force
    at finite temperature and $\mu=0$. In this figure, for
    $\kx=10^{-4}$ and $\kx=10^{-3}$, the jump from one to the other
    branch happens within the region enclosed by the dotted line.}
  \label{fig:nfksinf}
\end{figure}

At finite temperatures, there is a variation in the crosslink density
$n$ with the driving force $f$ (see Fig.~\ref{fig:nfksinf}). The
crosslinks unbind continuously with increasing force for soft
crosslinks (small $\kx$) while for large $\kx$, there is a jump in the
crosslink density indicating a first-order transition. The transition
is from a state where the filaments are tightly coupled by many bound
cross-links, to a state of nearly independent filaments with only a
few bound crosslinks. This happens due to the presence of a metastable
state at large $\kx$. The plot of the crosslink density with force,
therefore, shows a region where for a given force two values of $n$
are possible. The exact location of the jump in the crosslink density
from one branch to the other can be determined by observing when the
global minimum and the metastable minimum switch. This is the case for
slow quasi-static driving, when crosslinks are allowed to equilibrate
for a given level of the external load. Away from equilibrium, the
forward and the reverse branch will be different depending on the
driving frequency, giving rise to interesting dynamics and hysteresis
effects. This will be discussed later in Sect.\ref{sec:dyn}.
The range of $f$, when there is a metastable state, increases with
increasing crosslink stiffness. For very large crosslink stiffness,
the allowed values of $n$ are more towards the two extremes, $1$ and
$0$, like zero temperature behaviour. For a bundle consisting of more
than two filaments, a series of first-order transitions is expected as
observed in Ref.~\cite{vinkjcp}.

\begin{figure}[htbp]
  \centering
  \includegraphics[width=7cm,clip]{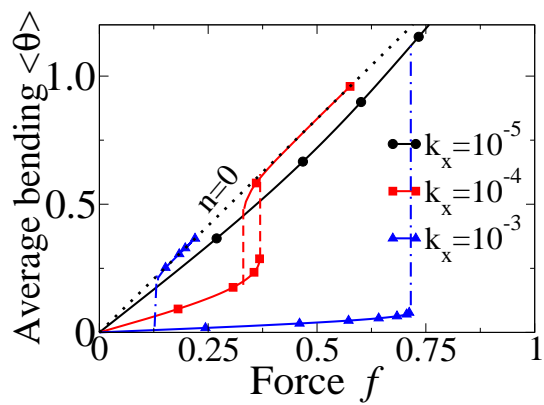}  
  \caption{This plot shows how the average bending $\ath$ of the
    bundle varies with the driving force $f$ ($\mu=0$). The black
    dotted line shows the behaviour for $n=0$. All the lines merge
    gradually to this dotted line for large force.}
  \label{fig:tfkxlinkksinf}
\end{figure}

The discontinuity in the crosslink density with force is reflected in
the average bending angle 
\begin{equation}
  \label{eq:avtheta}
  \ath=\frac{1}{L}\int_0^L\theta(s)ds,
\end{equation}
 which is feasible to observe in experiments. In
 Fig.\ref{fig:tfkxlinkksinf}, we plot $\ath$ as a function of force
 $f$. We see that the slope of the curve $\ath$ gradually reaches to a
 fixed value for all $\kx$, which corresponds to the value with
 $n=0$. For large $\kx$, the average bending changes suddenly from
 small value to a large one as a result of jump in the crosslink
 density. The softer the crosslinks, the smaller is the jump in the
 average bending at the critical force.
For the last two graphs, we set the chemical potential
$\mu=0$. Nonzero $\mu$ tends to lead to stronger discontinuities.

\begin{figure}[h]
  \centering
   \includegraphics[width=7cm,clip]{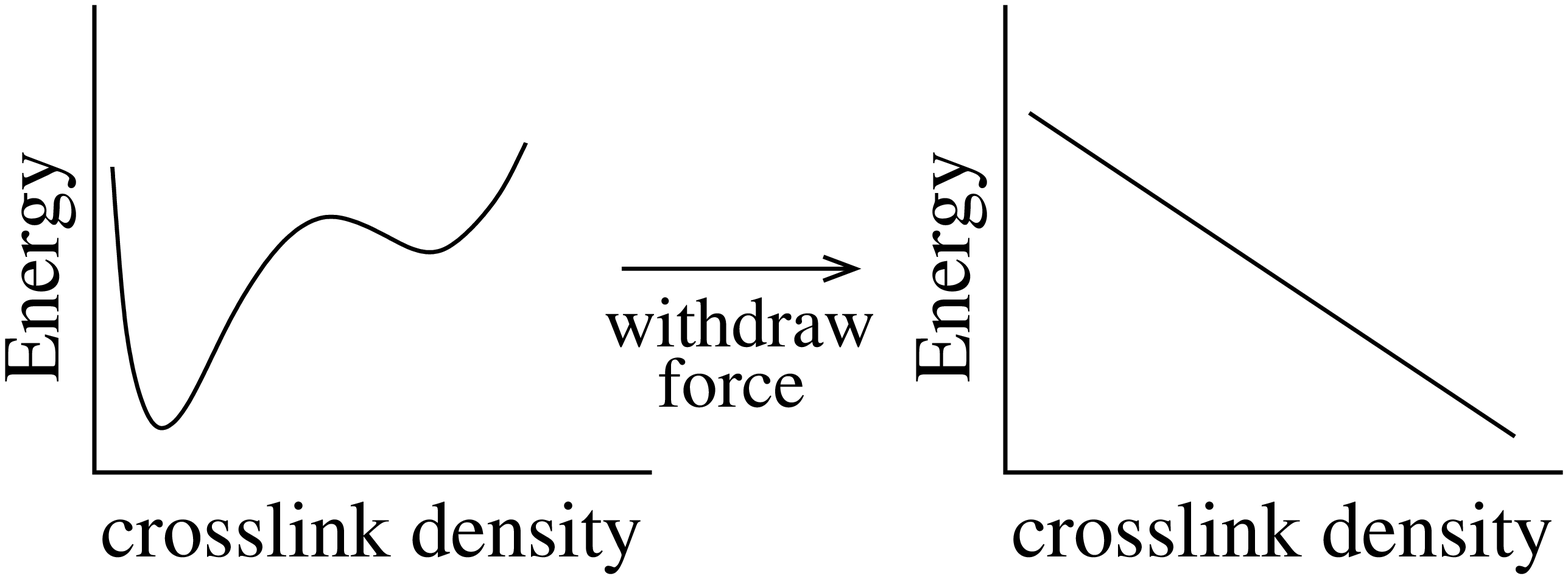}   
   \caption{Schematic of the energy landscape with and without
     force. At large forces, the bundle is in an unbound state with
     the minimum of the energy at very small $n$. At this condition,
     keeping all other parameters fixed, if we withdraw the force, the
     energy profile is linearly decreasing with $n$, with the minimum
     at $n=1$. Hence, after only a short time we see the system to
     roll back to the bound state, regaining its old configuration
     with perfectly ordered bound crosslinks as it was before the
     deformation.}
  \label{fig:Efunbind}
\end{figure}

This sudden unbinding of nearly all crosslinks is reminiscent of a
brittle failure process during which bundle integrity is lost
completely. With our choice of boundary conditions (grafted at one
end), filaments stay together, however, and after removal of the load,
crosslinks can again form. This will be a quick process which is
essentially downwards in energy landscape. This is illustrated by
Fig.~\ref{fig:Efunbind}.

A system with similar kind of unbinding mechanism is a double stranded
DNA. Force-induced phase transitions by pulling two strands in
opposite direction is discontinuous from a fully bound or zipped state
to a fully unbound or unzipped state~\cite{bhajpa}. This all-or-none
binding state resembles the case of the f-actin bundle with very stiff
crosslinks. 

\subsection{Rebinding of crosslinks under force}\label{sec:rebind}

\begin{figure}[htbp]
  \centering
  \includegraphics[width=7cm,clip]{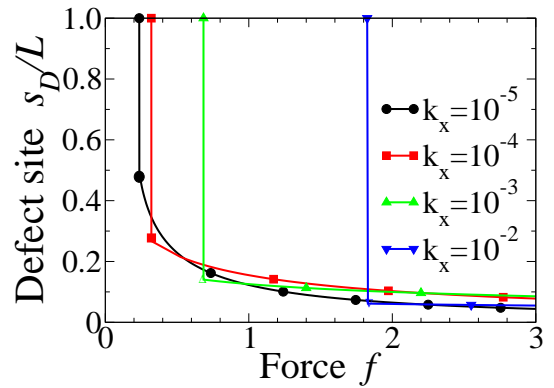}  
  \caption{Defect site $s_D$ moves with the driving force $f$. The
    critical force $f_c$ is the value of force where the lines jump to
    $s_D=L$. }
  \label{fig:sfksinf}
\end{figure}

In this section we consider the possibility that, under deformation,
crosslinks rebind to more favourable binding sites, thus forming a
defect. As explained in the modelling section this is accounted for by
a defect-function $d(s)$, which goes from zero to $\delta$ at the
defect site $s_D$. The relevant observable now becomes the defect site
$s_D$ instead of the crosslink density $n$, which we assume to be
saturated at $n=1$. The differential equation for $\theta(s)$,
obtained by minimizing the Hamiltonian w.r.t. $\theta$, is
\begin{equation}
  \label{eq:deksinf}
  -2\kf \theta''(s)+\frac{\kx b}{\delta}(b\theta(s)-d(s))=2f.
\end{equation}
One can find the total energy $E$ as a function of defect site $s_D$
using the solution $\theta(s)$ of this differential equation in the
Hamiltonian. The value of $s_D$ at which $E$ is lowest in the $E$ {\it
  vs.} $s_D$ curve gives the location of the defect. Without force,
$f=0$, it is at the free end, $s_D=L$. As the force increases, beyond
a critical force, the bundle creates a defect to reduce bending stress
and crosslink shearing. There is a region near the free end of the
bundle where the creation of a defect is very costly. This region
widens with increasing crosslink stiffness and vanishes for soft
crosslinks. Hence, for large $\kx$, when increasing force from a very
small value, we see a sudden creation of defect deep inside the bundle
at the critical force. Fig.~\ref{fig:sfksinf} shows the location of
the defect scaled by the length of the bundle $L$, {\it vs.} force
$f$. The critical force $f_c$ is the value of force where the defect
site jumps from $s_D=L$ to smaller value. As the force increases
further, the defect site moves towards the fixed end and gets stuck
there.

\begin{figure}[htbp]
  \centering
  \includegraphics[width=7cm,clip]{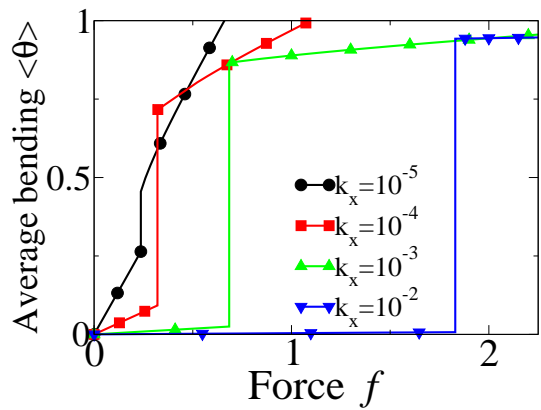}  
  \caption{The average bending $\ath$ as a function of driving force
    $f$ for various crosslink stiffness $\kx$.}
  \label{fig:tipksinf}
\end{figure}

The creation of the defect is also apparent in the value of the
average bending angle $\ath$, which displays a sudden jump. We plot
$\ath$ against force for different values of $\kx$ in
Fig.~\ref{fig:tipksinf}. For soft crosslinks the average bending angle
increases stronger with increasing force than that of stiff
crosslinks. Rather, for very stiff crosslinks, the average bending
does not change much with increasing force except at the critical
force, where we see a jump in the value of $\ath$. In this limit the
bundle is extremely stiff and can hardly be bent by the external force
-- any bending deformation would lead to very costly shearing of the
crosslinks, which is avoided as long as no defect forms. With a defect
present, the bend is localized to the defect region leaving most of
the bundle straight and thus the crosslinks unstrained.

\begin{figure}[htbp]
  \centering
  \includegraphics[width=6cm,clip]{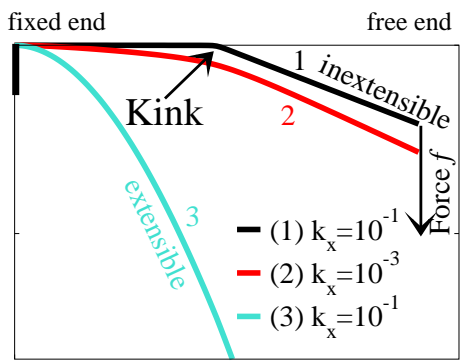}  
  \caption{Bundle contours for different crosslink stiffness. All the
    bundles have a defect at $s_D=L/2$. The sharpness of bending is
    much higher in the bundle with stiff crosslinks. The light
    coloured steep line (3) shows the bundle with extensible
    filaments.}
  \label{fig:contourksinf}
\end{figure}

This results in a kink in the bundle. Fig.\ref{fig:contourksinf} shows
how the bundle looks like for soft ($\kx=10^{-3}$) and stiff
crosslinks ($\kx=10^{-1}$) for a large force, $f=5f_b$. For
comparison, we fix the defect position at $s=L/2$ in both the
bundles. We see that for large $\kx$, there is a sharp bending, or
kink, around the defect site. The larger $\kx$, the sharper is the
kink. For small $\kx$ no kink is visible, even with a defect. The
light blue line in Fig.~\ref{fig:contourksinf} shows the bundle for
extensible filament with the same crosslink stiffness ($\kx=10^{-1}$)
as the kinked bundle (black line) with inextensible filaments. From
this graph we can conclude that the kink can be suppressed either by
allowing stretching in the filaments or by choosing soft crosslinks.

The stretching of the filaments will modify the shearing Hamiltonian by
an extra contribution of relative displacement $u(s)$ of the crosslink
binding sites of one filament with respect to the other in the
integrand, $b\theta(s)\to u+b\theta(s)$. Also there will be an extra
term in the Hamiltonian coming from the stretching energy of the
Hamiltonian, {\it viz.},
\begin{equation}
  \label{eq:Hst}
  H_{st}=\frac{k_s\delta}{2}\int_0^L u'^2(s)\, ds.
\end{equation}

\begin{figure}[htbp]
  \centering
  \includegraphics[width=7cm,clip]{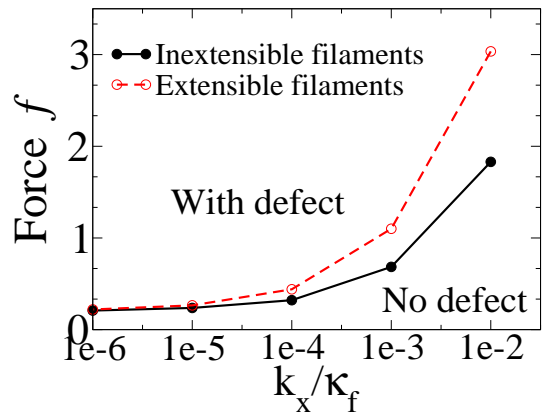}  
  \caption{The phase diagram of a bundle with inextensible filaments
    at zero temperature. The line separates the phases with a defect
    and with no defect. For small $\kx$, the critical force becomes
    independent of $\kx$. The dashed line shows the behaviour for
    extensible crosslinks.}
  \label{fig:phasediaksinf}
\end{figure}

The phase diagram in Fig.~\ref{fig:phasediaksinf} shows that the
critical force $f_c$ increases with increasing $\kx$. For very small
value of $\kx$, $f_c$ saturates to a fixed finite value, indicating
that no matter how soft the crosslinks are, there will be a
possibility to create a defect in the bundle by applying force.
In this graph, additionally, we indicate the phase separation line for
bundles with extensible filaments (dashed line). If we allow
stretching of the filaments, an additional deformation mode becomes
available that the bundle can use to minimize energy. Thus, stretching
delays the creation of a defect. So, for same crosslink stiffness, it
is easier to create a defect in bundles with inextensible filaments.

\begin{figure}[h]
  \centering
   \includegraphics[width=7cm,clip]{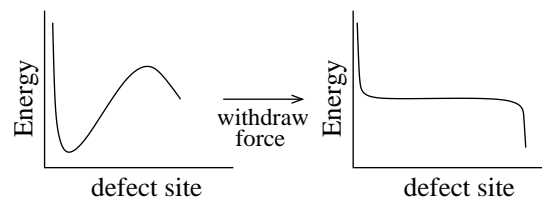}   
   \caption{Schematic of the energy landscape with and without
     force. At large forces, the bundle is in a kinked state with a
     defect present at small $s_D$. If we withdraw the force, the
     energy is flat and independent of $s_D$. This means that there is
     only a weak driving force for the defect to move towards the free
     end, where it can escape. Thus, depending on the dynamics of the
     defect, the bundle remains in its plastically deformed state for
     a long time.}
  \label{fig:Efrebind}
\end{figure}

Now let us consider a bundle which has been exposed to a large force
and held in the deformed configuration for a long time, so that there
is creation of a defect deep inside the bundle. In this configuration,
the energy $E(s_D)$ has a minimum at small $s_D$ with a large energy
barrier towards higher values of $s_D$ (see
Fig.~\ref{fig:Efrebind}). If, at this point, we remove the force, then
$E(s_D)$ has a wide plateau except at the two ends, the lowest energy
now being for a defect at the free end (i.e. no defect). Once the
defect is created in the bundle, and if the force is withdrawn, the
defect gets trapped in the plateau as it does not feel any driving
force to move towards the free end, where it can escape (see
Sect.~\ref{sec:coupl-dynam-evol}). Within this time scale, even
without any driving force, we see the bundle in a plastically deformed
configuration. This mechanism, which is reminiscent of the ductile
failure of metals, may very-well be responsible for the long-lived
residual deformation that has been observed in the experiments of
Strehle {\it et al.}~\cite{strehle}.

\section{Dynamics}\label{sec:dyn}

Let us now consider some aspects of the dynamical evolution of the
bundle degrees of freedom. In general, this will represent a coupled
evolution of the bundle contour, represented by $\theta(s)$, and the
binding state of the crosslinks, for example given in terms of the
average occupation $n$ or the defect location $s_D$. In what follows
we will present simplified approaches to deal with two scenarios of
time-dependent forcing: oscillating force, and switch-off after
constant force.

\subsection{Crosslink dynamics}

Here, we consider the brittle failure scenario of
Section~\ref{sec:unbind} and ask about the dynamical evolution of the
crosslink occupation $m=n\cdot N_\times=0,1,\ldots,N_\times$.

If bundle conformational degrees of freedom (bending mode $\theta$ and
possibly internal stretching modes) are assumed to be relaxed one can
formulate a one-step Master equation for the temporal evolution of the
probability distribution $p_m(t)$
\begin{equation}\label{eq:master.equation}
\frac{\partial p_m}{\partial t} = r_{m+1}p_{m+1} + g_{m-1}p_{m-1}-(r_m+g_m)p_m\,,
\end{equation}
with the rates
\begin{equation}
r_m = m e^{\beta\Delta E_m(f)}r_0\,, \qquad g_n = (N_\times-m)r_0\,.\nonumber
\end{equation}
$r_0$ represents an intrinsic rate constant and the free energy
profile $\Delta E_m(f)$ should be taken from
Section~\ref{sec:unbind}. A similar problem is the cluster of adhesion
sites discussed in Ref.~\cite{PhysRevLett.92.108102}. There, the
energy profile is taken to be linear in the applied force, which is
divided among all bound sites, $r_m = me^{f/m}r_0$. Here, the
dependence on force is, in general, more complex. It is also
illustrative to consider the general case in the framework of the
associated rate-equation
\begin{equation}\label{eq:rate.eqation}
\dot m = -mr_0e^{\beta\Delta E_m} +r_0 (N_\times-m)\,,
\end{equation}
When $\Delta E_m$ depends on $m$, this equation follows from the
Master equation Eq.~(\ref{eq:master.equation}) only by making the
approximation $\langle r_m\rangle\to r_{\langle m\rangle}$. As
expected from the results of Section~\ref{sec:unbind} (see
Fig.~\ref{fig:nfksinf}) this equation either has one stable stationary
state (at low or high force), or two stable states (at intermediate
force).

\begin{figure}[htbp]
  \centering
  \includegraphics[width=7cm,clip]{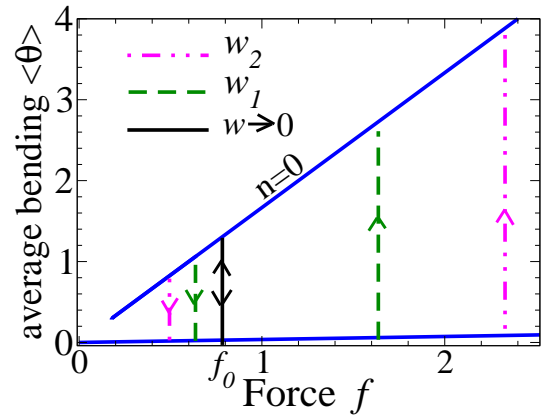}  
  \caption{Hysteresis loops for the average bending angle $\ath$
    vs. force $f$ for different frequencies ($\mu=0.01$). The black
    line corresponds to the small frequency limit, i.e. to the
    equilibrium transition. The green dashed line is for frequency
    $w/w_0=0.135$ ($w_1$) and the magenta dashed-dotted line
    corresponds to the frequency $w/w_0=0.368$ ($w_2$) . Larger
    frequency indicates larger hysteresis loop.}
  \label{fig:hystxlink}
\end{figure}

The full stochastic trajectory of the bundle occupation $m(t)$ in
response to time-dependent forces can only be obtained by solving the
Master equation Eq.~(\ref{eq:master.equation}). For oscillating forces
$f(t)=f_0\sin(\omega t)$, for example, one expects to see
$\omega$-dependent hysteresis effects, due to the presence of an
energy barrier between the metastable state and the groundstate. The
size of the energy barrier depends on force, $\Delta E(f)$, as
calculated in Section~\ref{sec:unbind}. With this dependence we can
set up a simplified treatment of barrier crossing events that lead to
hysteresis without having to solve the full Master equation. To this
end, we use Kramers equation for the rate $r$ of a thermally assisted
escape over an energy barrier
\begin{equation}
  \label{eq:fw}
  \Delta E(f)=-k_BT\ln \frac{r}{\hat r_0}\,,
\end{equation}
where $\hat r_0$ corresponds to some intrinsic attempt rate. By mapping
escape rate to frequency, $r/\hat r_0\equiv \omega/\omega_0$, we can
establish a relation $f(\omega)$, for the force at which the barrier
can be crossed. Fig.~\ref{fig:hystxlink} shows examples of the
resulting hysteresis loops for two different frequencies. As expected,
the area within the hysteresis loop is reduced when the frequency
decreases. The frequency range over which hysteresis can be observed
depends on the size of the energy barriers. For small values of the
chemical potential $\mu$, the barriers may only be a few $k_BT$. In
this limit our approach is not expected to hold.

\subsection{Coupled dynamical evolution of crosslinks and filament
  conformation}\label{sec:coupl-dynam-evol}

Here, we discuss the ductile failure mechanism introduced in
Section~\ref{sec:rebind}. In particular, we are interested in the
relaxation of the bundle contour after the force is switched off. This
situation is analogous to the experimental set-up of
Ref.~\cite{strehle}. The relaxation represents a coupled dynamical
evolution of the bundle contour $y(s,t)$ (with $y'\equiv\theta$) and
the defect location $s_D(t)$. For the contour one can derive the
equation of motion from the Hamiltonian Eqs.~(\ref{eq:Hb}) and
(\ref{eq:Hshd}).
\begin{equation}\label{eq:elasto.hydrodynamic}
\zeta \frac{\partial y}{\partial t} = -2\kappa_fy^{(4)} + \frac{\kx b}{\delta}\left ( by'' - d' \right )
\end{equation}
which represents the time-dependent generalization of Eq. (\ref{eq:deksinf}) for $f=0$.

For the temporal evolution of the defect position $s_D$ we use a
simple rate equation

\begin{equation}\label{eq:rate.defect}
\frac{\partial s_D}{\partial t} = r\delta\tanh(\beta\Delta E/2) 
\end{equation}
where the distance between crosslink binding sites $\delta$ represents
the length-scale for the defect motion; the rate $r$ sets the relevant
time-scale. It will depend, for example, on the chemical potential of
the crosslinks, $r\sim e^{\beta\mu}$.

\begin{figure}[htbp]
  \centering
  \includegraphics[width=7cm,clip]{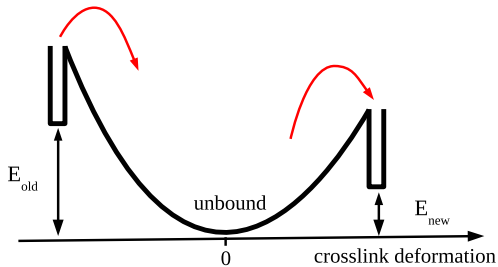}  
  \caption{Illustration of the energy landscape relevant for the
    motion of a defect. The defect moves by $\delta$, when one
    crosslink, first, unbinds and then rebinds to a neighboring
    site. The relative energy gain is $\Delta E = E_{\rm old}-E_{\rm new}$.}
  \label{fig:schematic.unbind.bind}
\end{figure}

The $\tanh$-factor derives from the activated nature of the
process. For the defect to move the distance $\delta$, two energy
barriers need to be crossed: first a crosslink has to unbind from its
old binding site. Second, it needs to ``stretch-out'' to reach its new
binding site. These two processes are illustrated in Fig.~\ref{fig:schematic.unbind.bind}.

The two Equations (\ref{eq:elasto.hydrodynamic}) and
~(\ref{eq:rate.defect}) have to be solved in parallel, with the defect
location entering the function $d$ in
Eq. (\ref{eq:elasto.hydrodynamic}), and the bundle contour determining
the energy gain $\Delta E$ in Eq. (\ref{eq:rate.defect}). This is
achieved via a mode-decomposition of the bundle contour
\begin{equation}
y(s,t) = y_{\rm part} + \sum_q \psi_q(s){\tilde y}_q(t)
\end{equation}
where the $\psi_q$ are the Eigenfunctions of the operator
$-\partial^4_s+\kappa_\times\partial^2_s$
~\footnote{The functions are
  given by $\psi_q=A\sin(qx)+B\cos(qx)+C\sinh(\tilde qx)+D\cosh(\tilde
  qx)$, where $\tilde q^2 = q^2 + \kappa_\times$ and the constants as
  well as the wavenumbers $q$ are determined from the boundary
  conditions.},
and we separate out a particular solution $y_{\rm part}$ of the
equation.  The resulting equations are integrated numerically with a
simple explicit Euler step.

\begin{figure}[htbp]
  \centering
  \includegraphics[width=7cm,clip]{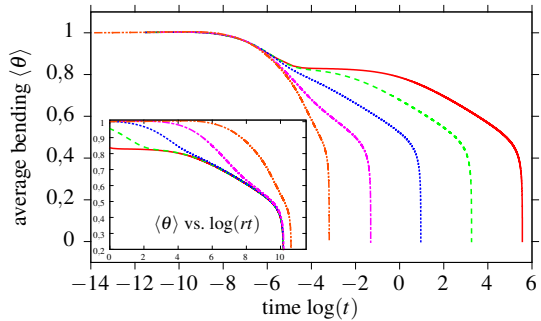}  
  \caption{Average bending angle $\langle\theta\rangle$ vs time $t$
    for different values of the time-scale
    $r^{-1}=10^{-6}\ldots10^{-2}$ (from left to right) and for
    $k_\times = 10^{-2}$.}
  \label{fig:theta.vs.time}
\end{figure}

Fig.~\ref{fig:theta.vs.time} displays the results of such a
calculation. Depicted is the average bending angle
$\langle\theta\rangle$ as a function of time and for various values of
the time-scale $r^{-1}$ for defect motion. The bundle is initialized
in a highly bent state with a defect at position $s_D=0.1$.

Before complete relaxation into its final straight state
($\langle\theta\rangle =0$), the bundle passes through different
stages. On small time-scales the evolution is independent of $r^{-1}$
and signals the relaxation of the bundle contour for an immobile
defect. Once this fast process is completed the final relaxation is
slaved to the motion of the defect. This is illustrated by the data
collapse when plotted vs. $rt$ as demonstrated in the inset. In the
plateau region the defect does not move as the driving force for motion
is very small ($\Delta E$ is small, see
Fig.~\ref{fig:Efrebind}). Depending on the time-scale for defect
motion and also on the crosslink stiffness $k_\times$, the bundle may
remain in this plastically deformed state for a long time. Finally,
the defect starts to move, leading to a logarithmically slow terminal
relaxation, after which the defect leaves the bundle.

\section{Summary and Outlook}\label{sec:summary}

In this paper, we study the response of F-actin bundles to driving
forces through a simple analytical model. We consider two filaments
connected by reversibly bound crosslinks and driven by a force applied
at the ends. Our model relies on the fact that under large
deformation, the crosslinks un- and re-bind in order to reduce the
stress in the bundle.

We can define two failure modes under load. \textit{Brittle failure} is
observed when crosslinks suddenly unbind, leading to catastrophic loss
of bundle integrity. \textit{Ductile failure} maintains bundle integrity
at the cost of crosslink remodelling and defect formation.

We present phase diagrams for the onset of failure, highlighting the
importance of crosslink stiffness for these processes. Crossing the
phase boundaries, force-deflection curves display
(frequency-dependent) hysteresis loops, reflecting the first-order
character of the failure processes.

We also relate our findings to recent experiments that evidence
long-lived plastically-deformed actin bundles~\cite{strehle}. To this
end we combine an elasto-hydrodynamic description of the bundle
relaxation with a rate equation for defect motion. The terminal
relaxation of the plastic deformation is thus seen to be slaved to the
dynamics of the defect. For weak crosslinks a defect is spread over
the entire length of the bundle and is not discernible in the bundle
contour. For stiff crosslinks defects take the form of well localized
kinks. These kinks only have a very small tendency to move towards the
free end of the bundle, where they can escape. Thus, crosslink
stiffness is a key factor governing the long-time dynamics of the
bundle.

The model should be extended towards a more realistic bundle
architecture, with more than just two filaments and in three spatial
dimensions. Similarly, it would be useful to consider more than one
defect and also different types of defects. Finally, it would be
interesting to analyze in more detail the transition between brittle
and ductile behavior, when, for example, the binding enthalpy of the
linkers is varied. These questions will have to be tackled with the
help of suitable simulation techniques and are left for future work.

\begin{acknowledgement}
  We acknowledge financial support by the DFG via the collaborative
  research center SFB 937, as well as via the Emmy Noether program (He
  6322/1-1).
\end{acknowledgement}


%

\end{document}